\documentclass[
 reprint,
 amsmath,amssymb,
 aps]{revtex4-2}

\usepackage{comment}
\usepackage{empheq}
\usepackage{amsmath}
\usepackage{bbold}
\usepackage{amssymb}
\usepackage{ulem,xpatch}
\usepackage{graphicx}
\usepackage{dcolumn}
\usepackage{bm}
\usepackage{listings}
\usepackage{siunitx}

\usepackage{appendix}

\newcommand{\hide}[1]{\relax}

\newcommand{\nocontentsline}[3]{}
\newcommand{\tocless}[2]{\bgroup\let\addcontentsline=\nocontentsline#1{#2}\egroup}

\begin{document}

\title{Membrane-in-the-middle optomechanics with a soft-clamped membrane at milliKelvin temperatures}%

\author{Eric Planz} 
    \thanks{These authors contributed equally.\\E. Planz now at Quantum Machines Denmark.}
    
\author{Xiang Xi}
    \thanks{These authors contributed equally.\\E. Planz now at Quantum Machines Denmark.}

\author{Thibault Capelle}
    \thanks{These authors contributed equally.\\E. Planz now at Quantum Machines Denmark.}

\author{Eric C. Langman}
\author{Albert Schliesser}
\email{albert.schliesser@nbi.ku.dk}
\affiliation{Niels Bohr Institute, University of Copenhagen, 2100 Copenhagen, Denmark}
\affiliation{Center for Hybrid Quantum Networks (Hy-Q), Niels Bohr Institute, University of Copenhagen, 2100 Copenhagen, Denmark}

\begin{abstract}
Soft-clamped silicon nitride membrane resonators are capable of coherence times $\tau$ exceeding 100 ms at millikelvin bath temperatures.
%
%
However, harnessing strong optomechanical coupling  in dry dilution refrigerators remains a challenge due to vibration issues and heating by optical absorption.
Here, we address these issues with an actuator-free optical cavity and mechanical resonator design, with the cavity mounted on a simple vibration-isolation platform.
We observe dynamical backaction when the cavity is driven with a free-space optical beam stabilized close to the red sideband using a two-beam locking scheme.
Finally, we characterize the effect of absorption heating on coherence time, finding it scales with the intracavity power $P$ as $\tau \propto P^{-\left( 0.34 \pm 0.04 \right)}$.
\end{abstract}

\maketitle

\section*{Introduction}

Cavity optomechanics has emerged as a dynamic field over the past few decades, fuelled by great progress in the fabrication of integrated, low-loss mechanical resonators \cite{Aspelmeyer2014}.
Coupling light to mechanical motion through radiation pressure effects in optomechanical systems has led to advances both in fundamental physics and technological applications.
Membrane-in-the-middle (MIM) systems, which utilize a partially reflective membrane resonator inside an optical cavity, have been of particular significance \cite{Thompson2008}.
They have been used in a wide variety of experiments, ranging from investigations of the tenets of continuous quantum measurement \cite{Purdy2013b, Moller2017, Rossi2018, Mason2019, Rossi2019}, to topological \cite{Xu2016} and parametric \cite{Haelg2022} energy transfer, and for applications in quantum information processing \cite{Brubaker2022}, gravitational wave detection \cite{Page2021}, and force sensing \cite{Fischer2019, Catalini2020}.

When operating MIM systems in the regime where motion of the mechanical oscillator is dominated by quantum uncertainties, it is necessary for the optomechanical coupling rates to exceed the thermal decoherence rate of the system.
Towards this goal, a large focus has been placed on developing mechanical resonators that demonstrate ultralow decoherence. 
%
%
\textit{Soft-clamped membrane resonators} \cite{Tsaturyan2017}, which comprise a phononic crystal pattern with an isolated defect, were used to reach the quantum regime at moderate ($T\sim 10$K) cryogenic temperatures \cite{Rossi2018}, as well as approach the quantum regime at room temperature \cite{Saarinen2023}.

%
The operation of such membranes in a dry dilution refrigerator is critical for applications such as electro-optic transduction\cite{Sahu2022, Brubaker2022} and realizing long-lived quantum memories\cite{Seis2022}. 
However, this introduces several challenges associated with (i) maintaining stability within the high-finesse cavity, and (ii) optical absorption heating of the membrane resonator at high intracavity fields.

%
%

Challenge (i) involves aligning to and locking the high-finesse cavity. Various approaches have been explored, including misaligning fiber-coupled cavities at room temperature to achieve high coupling efficiencies at low temperatures \cite{Fedoseev2022}.
To realize optical lock, actuators in fiber-coupled cavities \cite{Doeleman2023} and free-space coupling to optomechanical cavities \cite{Brubaker2022} have been used.
However, dry dilution refrigerators
tend to possess significant vibrations due to the use of a pulse tube system to maintain the Helium 4 at sufficiently low temperatures \cite{Olivieri2017}. These vibrations often result in large excursions that complicate locking to high-finesse cavities.

Challenge (ii) arises from the optical absorption induced heating of membrane resonators at high intracavity fields. This phenomenon, which has been observed in numerous optomechanical and electro-optic experiments \cite{Mirhosseini2020, Hease2020}, is problematic as it can lead to higher mechanical bath temperatures. Studies have shown that patterned SiN membranes demonstrate low thermal conductivity \cite{Leivo1998} and absorption heating effects can become significant at millikelvin temperatures, especially within the near-infrared regime \cite{Page2021}. Plain membranes with sub-millimeter dimensions have been employed in (wet) dilution refrigerators before, in which no significant heating was observed when driven with laser radiation at $\sim 1$-µm wavelenghth. (\cite{Peterson2016} made this observation by exposing a 100~nm thick membrane to 7.4~mW intracavity power, while \cite{Brubaker2022} report such a finding while shining 200~mW of intracavity power on a 40~nm thick membrane.)
Here, we investigate soft-clamped membranes resonators, which while offering higher quality factors ($Q>10^9$) and coherence times $\tau > 140 \text{ ms}$  \cite{Seis2022}, are expected to suffer from increased heating due to their patterned structure and larger dimensions ($\sim $cm). 

\begin{figure}[ht]
\centering
\includegraphics[width=1\linewidth]{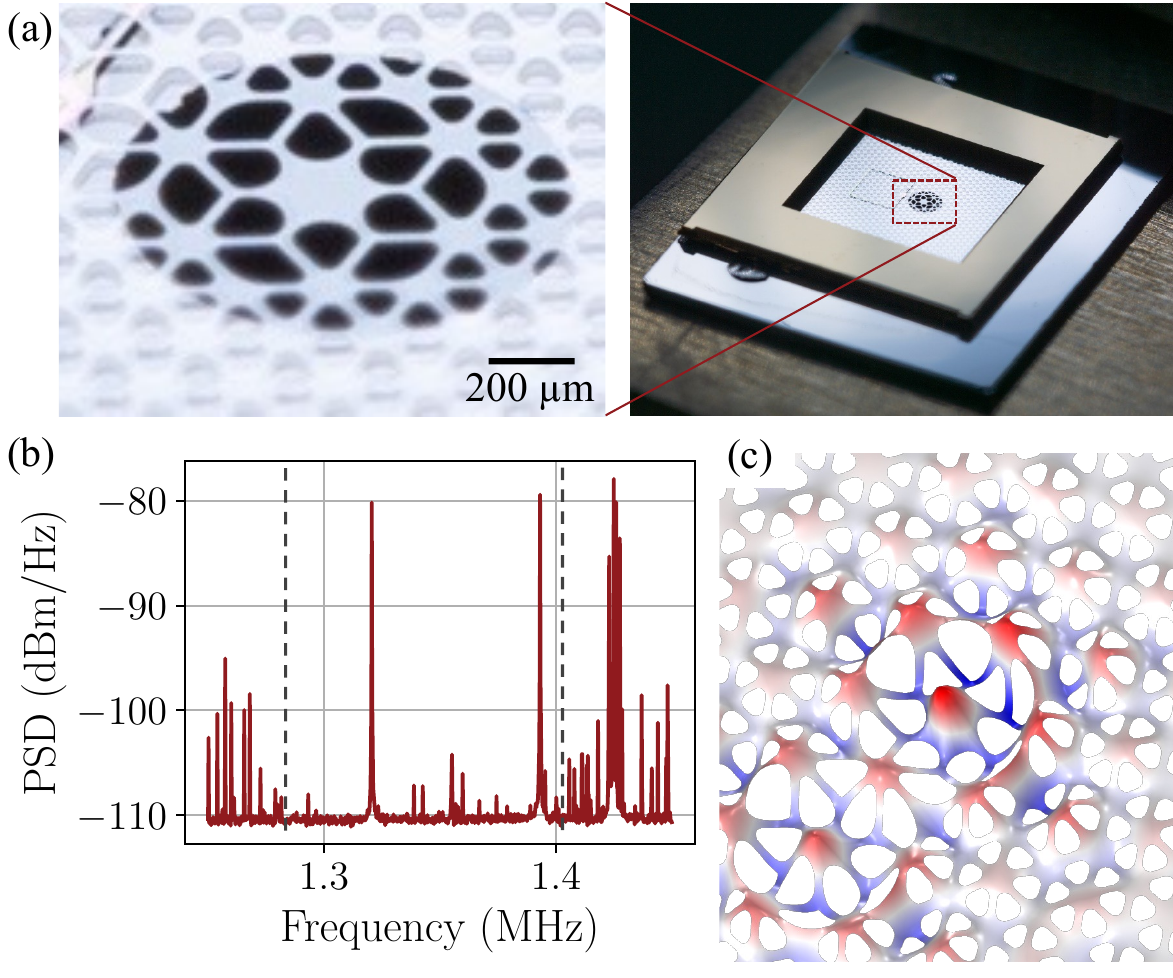}
\caption{(a) Photograph of the 'Lotus' membrane design used in the noise thermometry experiments. The defect diameter is designed to be 230 µm, corresponding to clipping losses low enough to achieve sideband-resolved optomechanics. (b) Room-temperature spectrum of the membrane showing a bandgap between the dashed lines and the defect mode at $1.32 \text{ MHz}$ that we address. (c) Simulated displacement of the mechanical mode localized at the defect in the phononic crystal patterned into a silicon nitride (SiN) membrane. Color map corresponds to displacement amplitude, from negative (blue) to positive (red).}
\label{fig:double_defect}
\end{figure}

In response to these issues, here we 
present a design for a sideband-resolved optomechanical assembly that offers a method for effective coupling and locking a laser to the cavity within a dilution refrigerator. 
Figure \ref{fig:double_defect} illustrates the soft-clamped membrane design with a mechanical resonance of interest at 1.32 MHz. The chip design incorporates additional coupling to a microwave cavity, although the details of this feature are outside the scope of this manuscript. We also investigate the absorption heating effect of the membrane from $805 \text{ nm}$ wavelength laser light.

\section*{Mechanical Design}

\begin{figure}[htb]
\centering
\includegraphics[width=1.06\linewidth]{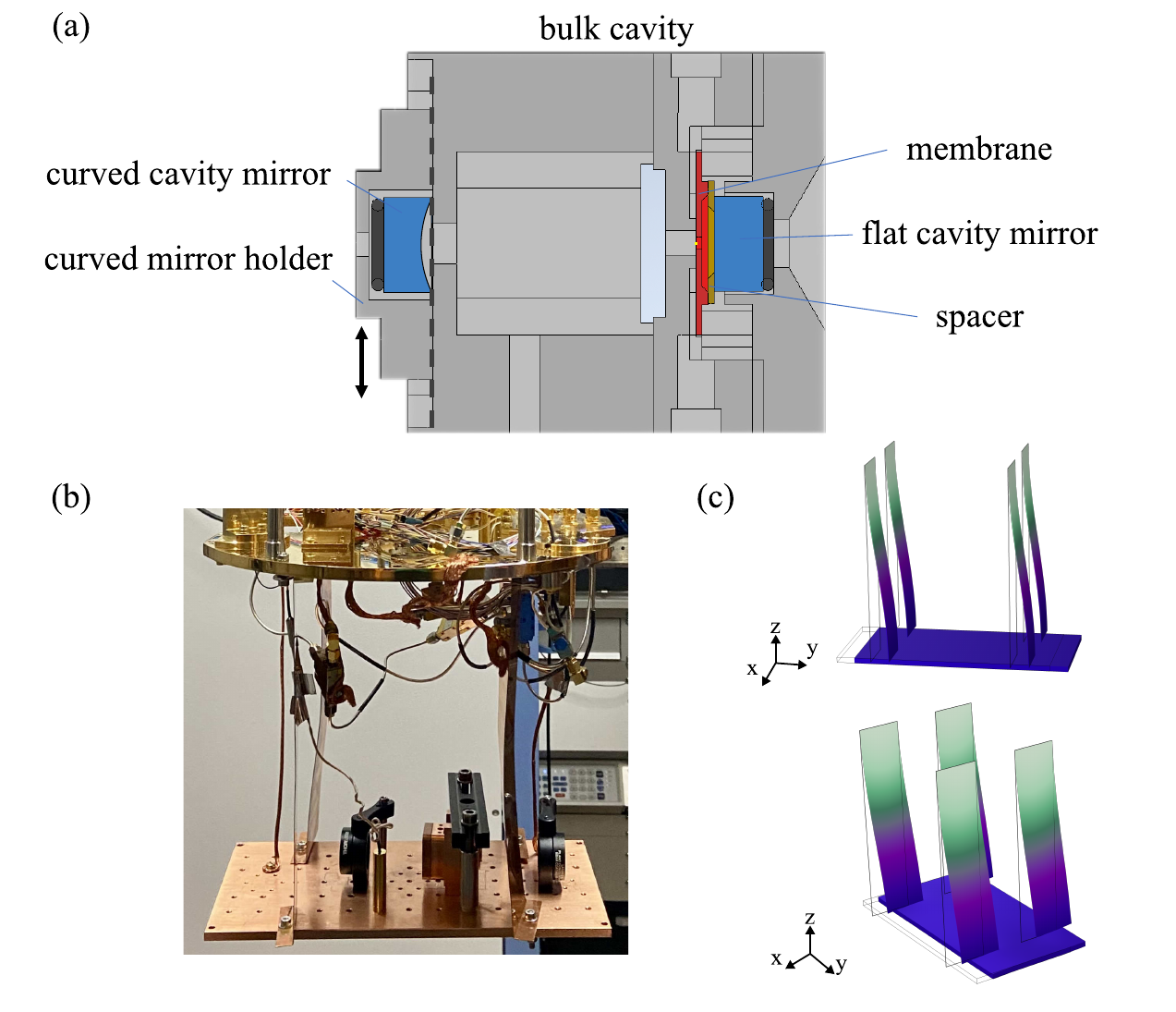}
\caption{(a) Schematic of the cavity assembly that shows the cavity mirrors (blue), the SiN membrane chip (red) and a silicon spacer chip (yellow), held in place by the copper parts of the assembly (gray).
In particular, two copper ``mirror holders'' press the two mirrors against the bulk cavity spacer using nitrile O-rings.
The curved mirror's holder can be moved orthogonal to the cavity axis, which allows centering the cavity field on the membrane defect. 
(b) Photograph of the vibration isolation platform mounted on the cryostat, with the cavity and the in/out coupling lenses.
(c) top (bottom): Displacement pattern of the lowest (second lowest)-frequency mode of the vibration isolation platform along the cavity axis (y-direction) (orthogonal to the cavity axis (x-direction)) at 2.3~Hz (145~Hz).}
\label{fig:Fig2-vibration}
\end{figure}

The cavity design shown in Fig. \ref{fig:Fig2-vibration}(a) employs a plano-convex, over-coupled Fabry-Pérot cavity with highly reflective mirrors. 
The wavelength-dependent reflectivities of these mirrors allows tuning both the finesse of the cavity and the over-coupling ratio by adjusting the wavelength. 
These features enable us to achieve a cavity finesse over 30{,}000, a cavity linewidth below 300~kHz, and substantial over-coupling greater than 95\%. 
The cavity assembly is made of oxygen-free high-conductivity copper.
%
The individual parts are clamped together tightly using short stainless steel screws to reduce fluctuations within the cavity due to differences in thermal contraction.
Within the assembly, the membrane frame lies parallel to the flat cavity mirror, separated by a 500~µm silicon spacer.
With an equally thick membrane chip, the membrane is located at approximately 1 mm from the flat mirror's surface \cite{Nielsen2016a}.
The total cavity length is $\sim 24 \text{ mm}$.
With the convex mirror's radius of curvature around $25 \text{ mm}$,  the waist of the optical mode at the position of the membrane is $\sim 43 \text{ µm}$.

The light is coupled to the cavity by aiming a free-space laser beam through windows in the cryostat, onto the more transmissive cavity mirror.
Movements of the cavity assembly in directions orthogonal to the longitudinal cavity axis can induce rather dramatic fluctuations in the intracavity field. Likewise, movements of the cavity in the axial direction can lead to a motion of the mechanical resonator not limited by thermal noise.
To address those two issues, the cavity assembly is affixed to the simple home-built vibration isolation platform shown in Fig. \ref{fig:Fig2-vibration}(b) and (c).
It consists of a heavy ($1.9 \text{ kg}$) rectangular (264 mm~$\times$~130 mm~$\times$~6.25 mm) copper plate that is suspended from the mixing chamber plate of a dry dilution refrigerator (LD250 by Bluefors) via thin copper sheets (24~cm$\times$5.1~cm~$\times$~0.6 mm), anchored at four points forming an inner rectangle of 170~mm~$\times$~104~mm.
This construction is `soft' for oscillations along the cavity axis, allowing the platform to swing at a low eigenfrequency of 2.3 Hz, with the aim of mitigating MHz mechanical noise due to non-thermal, external vibrations. For oscillations orthogonal to the cavity axis, it is much stiffer, with the first eigenfrequency appearing at 145 Hz in the horizontal direction. Here, the goal is to avoid large amplitude, low frequency excursions that lead to cavity axis pointing noise.
This platform allowed us to lock the laser to cavities with finesses up to $\sim31,000$, despite an active pulse tube disturbing the system (albeit without an intracavity membrane).  

\section*{Resonator Design}
\begin{figure*}[htb]
\centering
\includegraphics[width=1\linewidth]{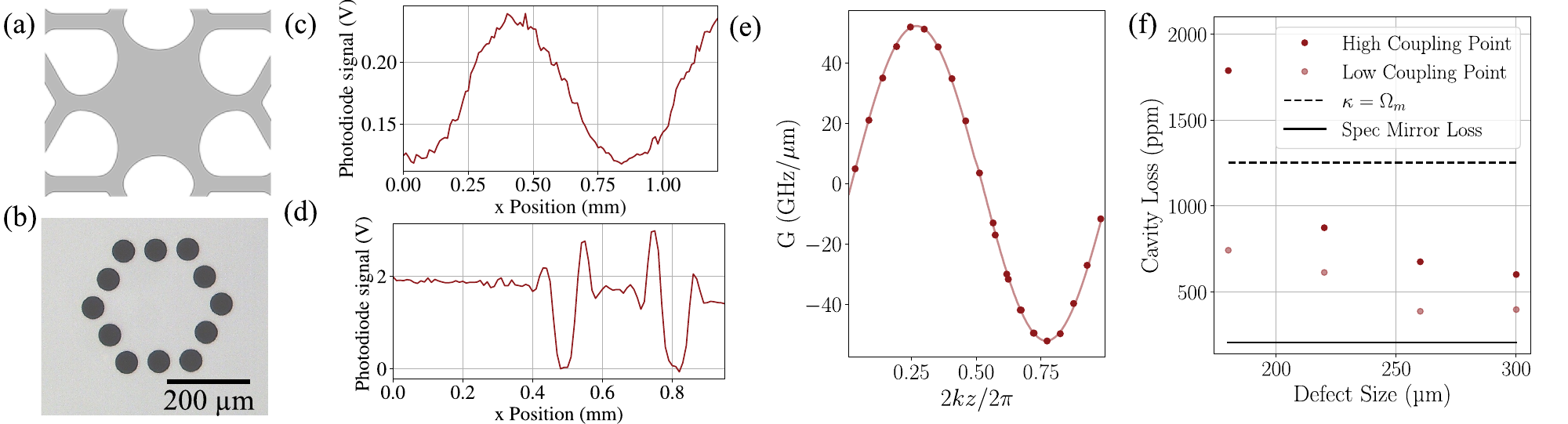}
\caption{
(a) Zoom-in to the geometry of a 'Lotus' defect. 
(b) Microscope picture of a 200~nm thick clipping test membrane used in the test assemblies.
The hexagonal pattern of holes simulates the hole structure in soft-clamped membranes. 
(c) Scan of the in-coupling laser beam along x-axis outside the hole structure with one cavity mirror removed, for a test membrane such as the one showed in (b). The back-reflected power shows interference between light being reflected off the membrane and the cavity mirror behind.
(d) Scan of the in-coupling laser beam after minimizing the tilt, for a test membrane such as the one showed in (b). The scan shows a clear feature stemming from the hole structure around the defect between 0.5 and 0.8~mm. These graphs are used to align the in-coupling laser beam to the center of the defect. 
(e) Optomechanical coupling at various wavelengths, as determined by measuring the frequencies of 24 subsequent optical resonances.
(f) Cavity loss seen in optomechanical cavity assemblies with various defect sizes. The data point at $\infty$ has been performed using a plane membrane, i.e., without hole structure.
}
\label{fig:Fig3-clipping}
\end{figure*}

This work utilizes a variant of Lotus-class soft-clamped membranes, which have demonstrated quality factors exceeding 1 billion in electromechanical experiments \cite{Seis2022}. Our experiment specifically employs a phononic dimer membrane containing two coupled defects leading to a pair of hybridized mechanical modes, one symmetric and the other antisymmetric \cite{Catalini2020}. The released membrane has a rectangular extent of $5.4 \text{ mm} \times 4.8 \text{ mm}$.
%
%
In our case, only one mechanical mode, at a frequency of $\Omega_m/2\pi\sim 1.32$~MHz will be considered. Fig. \ref{fig:Fig3-clipping}(a) shows a zoom-in of the mechanical defect of a 'Lotus' membrane described in Fig. \ref{fig:double_defect}. 

If the cavity mode diameter at the membrane position exceeds the defect size of a patterned membrane, additional optical losses occur. 
These `clipping losses' originate from the phase difference between the light field that travels through the material and that outside the defect. This results in altering the cavity wavefront and leads to coupling into higher-order modes. Minimizing these additional cavity losses is crucial to attain sideband resolution, 
a prerequisite for sideband cooling of a mechanical resonators to the quantum regime.
To gain insights into this cavity loss effect, we carried out a series of tests using optomechanical cavity assemblies with varied defect sizes. 
Figure \ref{fig:Fig3-clipping}(b) shows the hexagonal hole pattern etched into 200~nm thick SiN membranes, which we used in these tests to simulate the defect of a soft-clamped membrane.

In the assembly procedure for these tests, we first clamp the membrane and flat cavity mirror together.
Then, we scan a laser beam across the membrane, and measure the back-reflected power.
These scans typically show interference between the light reflected directly from the membrane and off the flat cavity mirror behind the membrane (Fig. \ref{fig:Fig3-clipping}(c)).
With 2D scans of the membrane plane, we can calculate the tilt between the membrane and cavity mirror from the interference fringe. 
In the example shown in Fig. \ref{fig:Fig3-clipping}(c), this leads to an angle of approx. $1 \text{ mrad}$ for one interference fringe across 0.8~mm with a 830~nm laser beam. 
For all test assemblies, the angle is kept $<$1~mrad to ensure tilt is not a dominant source of cavity loss. 

The same scan method is then used to find the center of the membrane defect with the laser beam.
In the example shown in Fig. \ref{fig:Fig3-clipping}(d), it occurs at a position around 7.93~mm. 
Once we have found the center position in the membrane plane, we position and fix the curved mirror in the assembly, forming the membrane-in-the-middle cavity.

The longitudinal position of the membrane with respect to the standing wave of the optical cavity determines the optomechanical coupling, among other factors.
To identify positions of high coupling, we measure a series of subsequent fundamental cavity resonance frequencies.
The position-dependent perturbation of the intracavity field through the dielectric SiN material causes a deviation of the cavity spectrum from equidistant modes separated by a constant free spectral range (FSR).
This deviation $\Delta \omega_\text{FSR}$ can be converted into a coupling point using the model provided in \cite{Dumont2019}.
It shows a periodicity of $2kz = 2\pi N z / L$, where $N$ is the mode number, $z$ the distance to the closest cavity mirror (here 1 mm), and $L$ is the cavity length (here 24~mm).
We therefore record the resonance frequency of 24 subsequent cavity modes, in order to observe all coupling points and obtain the graph in Fig. \ref{fig:Fig3-clipping}(e). 
Then we perform cavity loss measurements via cavity ringdowns at various coupling points, yielding the data presented in Fig. \ref{fig:Fig3-clipping}(f).
The findings demonstrate that the defect size does impact the cavity loss. 
We find that sufficiently small defects lead to significant enough cavity losses that sideband resolution becomes unattainable.
Based on these results, we use Lotus-class defects with an innermost diameter of 230~µm for relevant optical cavities.

\section*{Noise Thermometry}

\begin{figure}[htbp]
\centering
\includegraphics[width=1\linewidth]{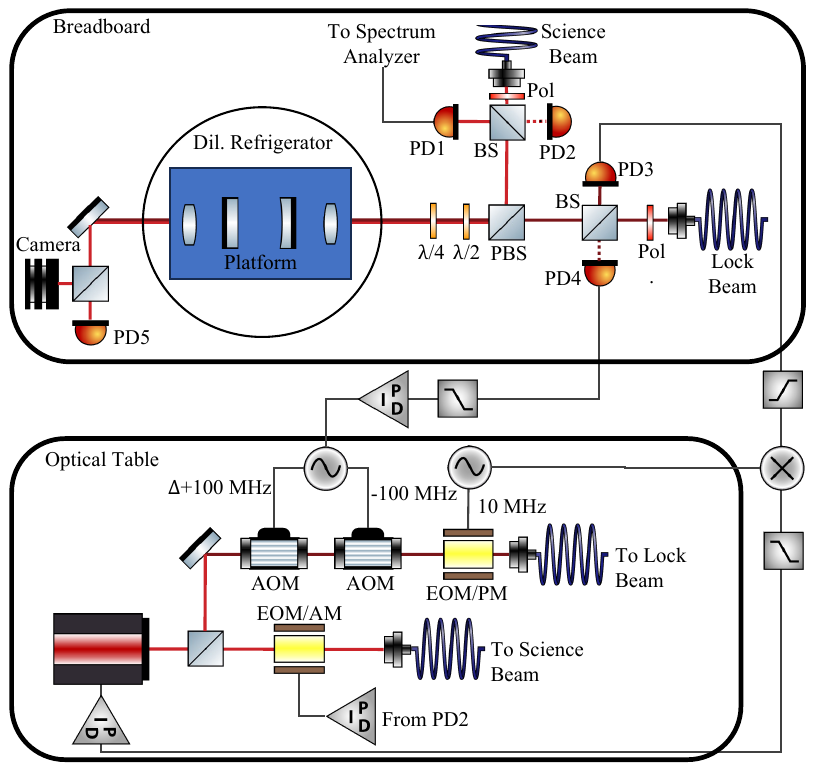}
\caption{Experimental Setup.
Optical lock and science beams are derived from the same laser, frequency-offset by two acousto-optic modulators.
The two beams are prepared in orthogonal polarization with linear polarizers (Pol.), and subsequently combined using a polarizing beam splitter (PBS).
In a combination with a set of waveplates ($\lambda/4$,$\lambda/2$), the same PBS separates the beams reflected from the cavity again.
Non-polarizing beam splitters (BS) reroute fractions of the lock and science input powers towards photodetectors PD4 and PD2, respectively.
Their signals are used to stabilize the two beams' powers by feeding back to an electrooptic amplitude modulator (EOM/AM) and to the power of the radio-frequency signal driving the AOMs, respectively.
The laser frequency is locked to the cavity via an error signal generated by demodulating the photocurrent detected on photodetector PD3, with feedback to a piezoelectric actuator that adjusts the frequency of the laser.
The optomechanical cavity is mounted on the vibration isolation platform inside the dilution refrigerator together with two in- and out-coupling lenses. 
The photodetector PD1 records the intensity of the detuned science beam. 
The photocurrent spectra, as recorded with an R\&S FSW26 spectrum analyser, therefore also contain the thermomechanical noise of the membrane resonances.
Additionally, the cavity mode can be monitored in transmission via a camera and PD5.
}
\label{fig:optics_thermometry}
\end{figure}

For the optomechanical measurements at millikelvin temperatures, we clamp the cavity assembly to the vibration isolation platform described in Fig. \ref{fig:Fig2-vibration}(b).
We shine laser light towards the cavity, through the cryostat's windows, from fiber couplers mounted on a  breadboard, which is itself clamped to the outer shield of the dilution refrigerator. 
Figure \ref{fig:optics_thermometry} shows a simplified version of the optical setup, which uses two orthogonally polarized beams derived from the same laser via two acousto-optic modulators (AOMs) \cite{Brubaker2022}.
In our experiments we use a widely tuneable, low-noise Ti:sapphire laser SolsTiS from MSquared at a wavelength around 800~nm.
The first, weak ``lock'' beam is parked at resonance with the cavity and its reflection is used to derive a Pound-Drever-Hall (PDH) error signal to lock the laser frequency to the cavity.
The second, stronger, red-detuned, ``science'' beam probes (and cools) the mechanical resonance.
Its detuning from the cavity resonance is set by the frequencies at which the two free-space AOMs from CSRayzer are driven.
We record mechanical noise spectra by direct detection of the science beam using avalanche photodiodes of type APD410A2 from Thorlabs.

\begin{figure}[htb]
\centering
\includegraphics[width=0.95\linewidth]{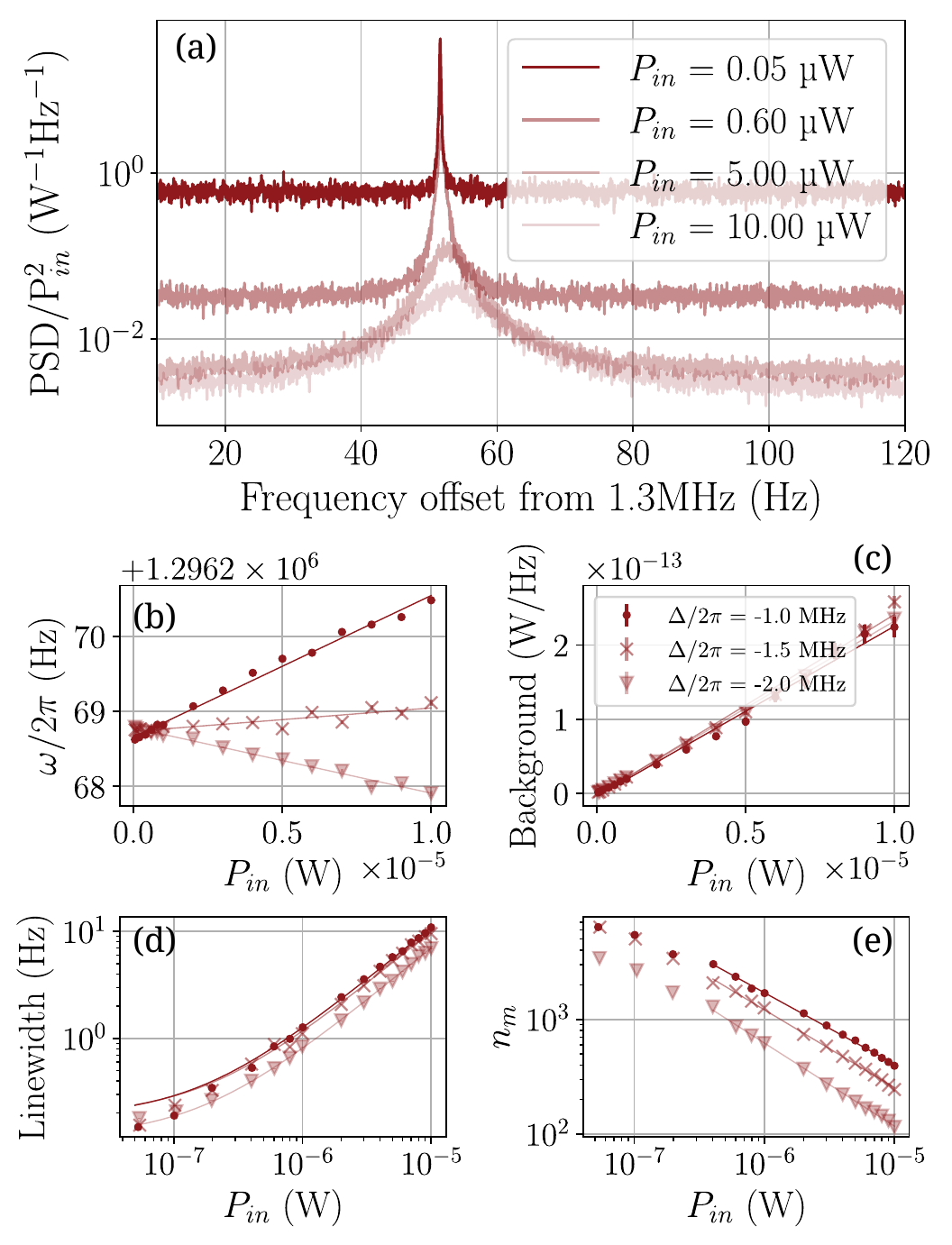}
\caption{
Noise Thermometry. 
(a) Mechanical spectrum PSD normalized to the input optical power squared for different optical input powers showing cooling of the mechanical resonance. For example, $P_\text{in} = 5 \text{ µW}$ at $-1.5 \text{ MHz}$ detuning corresponds to an intracavity photon number of $2.2 \times 10^6$.
(b) Shift in the mechanical frequency at various optical powers. The different symbols, for the Fig 5(b) to 5(e), correspond to different detunings, as specified in the legend of Fig. 5(c).
(c) Noise spectrum background scaling linearly, thereby indicating a shot-noise limited measurement. (d) Mechanical linewidth at various powers showing dynamical backaction. 
(e) Phonon occupation during cooling power sweeps indicating power-law scaling. The normalization is obtained from the so-called "Gorodetsky method"\cite{Gorodetksy2010} during a separate measurement on the same sample at first high (>$\approx$4 K) baseplate temperature to obtain an estimate of $g_0$, then at low ($\approx 15$~mK) baseplate temperature using this estimate of $g_0$ to retrieve an estimate of $T_\mathrm{bath}=643\pm 108$~mK. The small dynamical backaction in the low powers of the shown data set is taken into account, but not the small amount of optical heating expected here, which is why the real mechanical occupancy is expected to be slightly higher than shown.}
\label{fig:thermometry}
\end{figure}

We initially measure the thermal noise spectrum at room temperature, revealing the fundamental mode of the soft-clamped membrane at approximately 1.32~MHz (see Fig. \ref{fig:double_defect}(b)). 
As the setup cools, thermal contraction of the silicon frame triggers an approx. 2\% down-shift in mechanical frequency. 
At millikelvin temperatures we observe the same resonance close to 1.30~MHz.
As we increase the science beam power, we observe dynamical backaction damping as a broadening mechanical linewidth \cite{Aspelmeyer2014}, as shown in Fig.~\ref{fig:thermometry}(a).

For a systematic laser cooling series, we maintain the dilution refrigerator temperature at 20~mK and perform measurements at the science beam detunings [-1.0, -1.5, -2.0]~MHz. 
The science beam's input power is swept up to $10 \text{ µW}$, at an estimated mode-matching efficiency of $0{.}8$.
A fit to the PDH error signal gives a cavity linewidth $\kappa / 2 \pi = \left( 2.0 \pm 0.2 \right) \text{ MHz}$, corresponding to a finesse of $\left( 3.1 \pm 0.3 \right) \times 10^3$.

Figures \ref{fig:thermometry}(b-d) show the results of Lorentzian fits to these spectra. 
We note resonance frequency shifts of different signs for distinct detunings, attributed to the anticipated optical spring effect.
%
%
Moreover, the linewidth increase directly relates to the dynamical backaction effect, from which we infer a vacuum optomechanical coupling rate \cite{Aspelmeyer2014} of $g_0 / 2 \pi \approx  1.2 \text{ Hz}$. For comparison, the maximum achievable coupling rate in a membrane-in-the-middle configuration can be approximated as
$ g_0 ^\text{max} \approx 2 (\omega_\text{c}/{L}) |r| x_\text{zpf} \xi $,
where $\omega_\text{c}/2\pi$ is the cavity resonance frequency, $L$ is the cavity length, $r$ the optical field reflectivity of the membrane, $x_\text{zpf}$ is the zero-point fluctuation, and $\xi$ is the mode overlap between membrane displacement and optical field \cite{Saarinen2023}. With the $50 \text{ nm}$ thick membrane and $24 \text{ mm}$ long optical cavity, we anticipate $g_0 ^\text{max} / 2 \pi \approx 8 \text{ Hz}$ at perfect mode-overlap.
We attribute the discrepancy with the measured $g_0$ to the unoptimized positioning of the membrane along the cavity axis and a potentially imperfect transverse overlap ($\xi<1$).

Finally, in Fig. \ref{fig:thermometry}(e) we plot the area $A$ of the mechanical peak over the intracavity power $P$ squared.
In a direct-detection measurement, as in this setup, $A/P^2$ is proportional to the steady-state phonon occupation number $\bar{n}_\text{f}$. 
In our experiment's regime, quantum backaction is negligible and the occupation is approximately given by
\begin{align}
    \bar{n}_\text{f} = \frac{\Gamma_\text{m} \bar{n}_\text{th}}{\Gamma_\text{opt} + \Gamma_\text{m}} \approx \frac{\Gamma_\text{m} \bar{n}_\text{th}}{\Gamma_\text{opt}} \text{ ,}
    \label{eq:nf}
\end{align}
where $\Gamma_\text{m}$ is the mechanical damping rate, $\bar{n}_\text{th}$ is the occupation of the mechanical bath, and $\Gamma_\text{opt}$ is the optomechanical damping rate \cite{Aspelmeyer2014}.
As $\Gamma_\text{opt}\propto P$ \cite{Aspelmeyer2014}, we would expect a power law $A/P^2\propto \bar n_\mathrm{f}\propto P^s \propto n_\mathrm{cav}^s$ with slope $s=-1$ to emerge in the log-log-scale in Fig. \ref{fig:thermometry}(e)---were there no additional heating effects. 
However, we obtain a slope of $s=-0.67 \pm 0.04$.
Combined with eq.~(\ref{eq:nf}), this suggests a power scaling of the decoherence rate $\Gamma_\text{m} \bar{n}_\text{th}\propto P^\alpha$ with $\alpha = (-1-s) = 0.33 \pm 0.04$, and consequently the (heating-limited) coherence time $\tau\propto P^{-\alpha}$.
This is in rough agreement with published results of the optical absorption heating and mechanical linewidth broadening of soft-clamped membranes.
Indeed, reference \cite{Page2021} (Fig. S2) suggests that the mechanical temperature scales approximately as:
\begin{equation}
\label{OPAB}
\bar{n}_\text{th}\propto P^{0.33},
\end{equation}
whereas the articles \cite{Seis2022} and \cite{Zhou2019} find a scaling of the damping with the mechanical bath temperature close to:
\begin{equation}
\label{Gammam_vs_T}
\Gamma_\text{m}\propto n_\mathrm{th}^{0{.}66}.
\end{equation} 
By combining Eqs. \ref{OPAB} and \ref{Gammam_vs_T}, a crude approximation would then suggest a scaling:
\begin{equation}
\Gamma_\text{m} \cdot \bar{n}_\text{th}\propto (P^{0.33})^{0.66}\cdot P^{0.33}\propto P^{0.55},
\end{equation}
that is $\alpha_\text{lit.} \approx 0.55$, which has to be compared to our findings of $\alpha\approx 0.33$. We attribute the small discrepancy between those values to the questionable extrapolation we use to combine Eqs. (\ref{OPAB}) and (\ref{Gammam_vs_T}), which have been measured in very different experimental situations.

\section*{Conclusion}
We have presented a detailed experimental procedure to construct and analyze an optomechanical setup with a Lotus-class soft-clamped membrane within an over-coupled Fabry-Pérot cavity.
The setup operates in a dry dilution refrigeration thanks to a stable mechanical design and a vibration isolation platform. 
We have explored how the finite size of the membrane defect can introduce additional cavity losses.
We have developed an efficient procedure of assembling MIM cavities and derived design restrictions on soft-clamped membranes, here being a defect size $>200 \text{ µm}$.
The thermal noise spectra observed during the noise thermometry experiments shows clear dynamical backaction cooling.
From the noise thermometry measurement, we derived the power-law scaling of the decoherence rate with the optical power with an exponent of $\alpha = 0.33 \pm 0.04$, in reasonable agreement with existing literature.
This suggests that soft-clamped membranes are  subject to optical absorption heating and mechanical linewidth broadening.

Our findings provide insights for researchers in quantum optomechanics to optimize their experimental procedures. 
The techniques and approaches developed here could facilitate the realization of new optomechanical systems that operate in the quantum regime. 
In the future, coupling the second defect of the phononic dimer to a microwave resonator could enable quantum transduction experiments, with long-term intermediate storage of the quantum state, opening new possibilities in quantum communication and computation.
Other applications include compact setups for manipulating optical quantum noise \cite{Qin2014, Page2021}, as well as the search for unconventinoal decoherence \cite{Nimmrichter2014} and dark matter \cite{Carney2021}.

\section*{Acknowledgements}
This work was supported by the European Research Council project PHOQS  (grant no.~101002179), the Novo Nordisk Foundation (grant no.~NNF20OC0061866) and the Danish National Research Foundation (Center of Excellence “Hy-Q”).
This project has furthermore received funding from the European Union's Horizon 2020 research and innovation programme under the Marie Skłodowska-Curie grant agreements No. 801199 and 101107341.\\
\\
\textbf{Disclosures.}
The authors declare no conflicts of interest.\\
\\
\textbf{Data availability.}
Data underlying the results presented in this paper are available on \url{https://doi.org/10.5281/zenodo.8207950}.

\bibliography{refs}

\end{document}